\documentstyle[12pt]{article}
\newcommand{\be}[1]{\begin{equation} \label{(#1)}}
\newcommand{\ee}{\end{equation}}
\newcommand{\ba}[1]{\begin{eqnarray} \label{(#1)}}
\newcommand{\ea}{\end{eqnarray}}
\newcommand{\nn}{\nonumber}
\newcommand{\rf}[1]{(\ref{(#1)})}

\def\rp{$R_p \hspace{-1em}/\;\:$}
\def\rpm{R_p \hspace{-0.8em}/\;\:}

\def\pmb#1{\setbox0=\hbox{#1}%
  \kern-.015em\copy0\kern-\wd0
  \kern.03em\copy0\kern-\wd0
  \kern-.015em\raise.0233em\box0 }
\def \znbb {0\nu\beta\beta}

\def\bfr{\pmb{${r}$}}
\def\bfsgm{\pmb{${\sigma}$}}
\def\sir{({ \bfsgm_{i}^{~}} \cdot {\hat{\bfr}_{ij}^{~}} )}
\def\sjr{({ \bfsgm_{j}^{~}} \cdot {\hat{\bfr}_{ij}^{~}} )}
\def\si{{ \bfsgm_{i}^{~}}}
\def\sj{{ \bfsgm_{j}^{~}} }
\begin{document}

\begin{center}

{\bf New Supersymmetric  Contributions to Neutrinoless Double Beta
Decay}

\bigskip

{M. Hirsch\footnotemark[1], H.V. Klapdor-Kleingrothaus\footnotemark[2]
\bigskip

{\it
Max-Plank-Institut f\"{u}r Kernphysik, P.O. 10 39 80, D-69029,
Heidelberg, Germany}

\bigskip
S.G. Kovalenko\footnotemark[3]
\bigskip

{\it Joint Institute for Nuclear Research, Dubna, Russia}
}

\end{center}

\begin{abstract}

The neutrinoless double beta ($\znbb$) decay is analyzed
within the Minimal Supersymmetric Standard Model with explicit
R-parity violation (\rp  MSSM).
We have found new supersymmetric contributions
to this process and give the complete set of relevant
Feynman diagrams. Operators describing $0^+ \longrightarrow 0^+$
nuclear transitions induced by the supersymmetric interactions
of the \rp MSSM are derived. These operators can be used
for calculating the $\znbb$ decay rate applying any specific nuclear
model wave functions.
\end{abstract}

\bigskip
\footnotetext[1]{MAHIRSCH@ENULL.MPI-HD.MPG.DE}
\footnotetext[2]{KLAPDOR@ENULL.MPI-HD.MPG.DE}
\footnotetext[3]{KOVALEN@NUSUN.JINR.DUBNA.SU}

The observation of neutrinoless double beta ($\znbb$) decay
would be a clear signal for physics beyond the standard
model, since it violates lepton number by two units
(for reviews see \cite{hax84}, \cite{doi85}). No definite
observation of $\znbb$ decay has been reported to date, but
recent experimental progress has pushed the existing
half-life limits of $\znbb$ decay beyond $10^{24}$ years and
further progress can be expected in the near future \cite{hdmo94}.

Half-life limits on $\znbb$ decay are usually interpreted as
limits on the effective Majorana neutrino mass (see fig. 1). However,
it is known for some time that
there exist also other mechanisms which might induce $\znbb$ decays,
like for example in left-right symmetric extensions of the standard
model gauge group \cite{doi85}.
In this paper we study contributions to $\znbb$ decay
within supersymmetric (SUSY) theories with explicit R-parity breaking.
R-parity ($R_p$) is a discrete, multiplicative symmetry
defined as $R_p = (-1)^{3B+L+2S}$, where $S,\ B$ and $L$
are the spin, the baryon and the lepton quantum number.
Conservation of R-parity in turn implies baryon number ($B$)
and lepton number ($L$) conservation. This symmetry has been imposed on
the minimal supersymmetric standard model (MSSM) (for a review
see \cite{Haber}). However, neither gauge invariance nor
supersymmetry require $R_p$ conservation.
The question whether or not $R_p$ is a good symmetry of
the supersymmetric theory is a dynamical problem which might be
related to more fundamental physics at the Planck scale.
In general, $R_p$ can be either broken explicitly
{}~\cite{dh88}, ~\cite{hs84} or spontaneously ~\cite{R_spont},
{}~\cite{mv90} by the vacuum expectation value of the scalar superpartner
of the $R_p$-odd  isosinglet lepton field ~\cite{mv90}.
Supersymmetric models with $R_p$ non-conservation (\rp) have been
extensively discussed in the literature not only because of their great
theoretical interest, but also because they have interesting phenomenological
\cite{bgh}, \cite{zwirn83}, ~\cite{weinberg82},
and cosmological \cite{Cosmology}-\cite{dr94} implications.
One can expect that \rp SUSY models give a very natural framework
for lepton number violating
processes and particularly $\znbb$ decay ~\cite{Mohapatra}.

Let us start with a brief review of the model we use in the present paper.

We follow the MSSM extended by inclusion of explicit $R$-parity
violating (\rp) terms  into the superpotential .
This model (\rp MSSM)  has the MSSM field content and
is completely specified
by the standard $SU(3)\times SU(2)\times U(1)$ gauge couplings
as well as by the low-energy superpotential and "soft" SUSY
breaking terms \cite{Haber}.
The most general gauge invariant form of the superpotential is
{}~\cite{weinberg82},~\cite{drw}
\be{sup_gen}
           W = W_{R_p} + W_{\rpm}.
\ee
The $R_p$ conserving part has the standard MSSM form
\be{R_p-cons}
         W_{R_p} = h_L H_1 L {\bar E} + h_D H_1 Q {\bar D} -
         h_U H_2 Q {\bar U} -
	\mu H_1 H_2.
\ee
We use notations $L$,  $Q$ for lepton and quark
doublet superfields and ${\bar E}, \ {\bar U},\  {\bar D}$ for lepton and
{\em up}, {\em down} quark singlet  superfields;
$H_1$ and $H_2$ are the Higgs doublet superfields
with a weak hypercharge $Y=-1, \ +1$, respectively.
Summation over generations is implied.
For simplicity generation indices of fields and Yukawa coupling
constants  $h_L, \ h_U, \ h_D$  are suppressed.
The mass-mixing parameter $\mu$ is a free parameter describing
mixing between the Higgs bosons $H_{1}$-$H_{2}$
as well as between higgsinos $\tilde H_1$-$\tilde H_2$.

The $R_p$ violating part of the superpotential \rf{sup_gen}
can be written as ~\cite{dh88}, ~\cite{hs84},
\be{R-viol}
W_{\rpm} = \lambda_{ijk}L_i L_j {\bar E}_k + \lambda'_{ijk}L_i Q_j {\bar D}_k
+ \lambda''_{ijk}{\bar U}_i {\bar D}_j {\bar D}_k
\ee
where indices $i,j, k$ denote generations, and the fields have been defined
so that the bilinear lepton number violating operators  $L_i H_2$ ~\cite{hs84}
have been rotated away.
The coupling constants $\lambda$ ($\lambda''$) are antisymmetric in the first
(last) two indices. The first two terms lead to lepton number violation,
while the last one violates baryon number conservation.
Proton stability forbids simultaneous presence of lepton and baryon number
violating terms in the superpotential \cite{zwirn83} (unless the couplings
are very small).  Therefore, only $\lambda$, $\lambda'$ {\it or} $\lambda''$
type interactions can be present. There may exist an underlying
discrete symmetry in the theory which allows either the first or
the second set of couplings.
This discrete symmetry, called matter parity ~\cite{weinberg82},
{}~\cite{drw} can be imposed 'ad hoc' on the MSSM.
It can be justified on a more fundamental level of Planck scale physics.
Such a discrete symmetry is shown to be compatible with the ordinary
$SU(5)$ ~\cite{hs84} and "flipped" $SU(5)\times U(1)$ ~\cite{bh89}
grand unification (GUT) scenarios, as well as with phenomenologically viable
superstring theories ~\cite{sstr}.

Neutrinoless double beta decay, which is the main subject of the present paper,
requires lepton number violating interactions.
Therefore we bind ourselves to the \rp MSSM with
lepton number violation ($\lambda\neq 0,\   \lambda'\neq 0$)
and baryon number conservation ($\lambda'' = 0$).
 Apparently, $\znbb$ can probe only the first generation
lepton number violating Yukawa coupling $\lambda'_{111}$ because only
the first generation fermions $u, d, e$ are involved in this process.

In addition to proton decay constraints on \rp couplings
there are also constraints which follow from cosmological arguments
requiring that the baryon asymmetry generated at the GUT scale is not
washed out by $B-L$ violating interactions present in eq. \rf{R-viol}.
These cosmological constraints have been thought to affect all \rp couplings
$\lambda, \lambda', \lambda''\ll 10^{-7}$, making
these models phenomenologically not interesting.
These arguments, however, were proved to be strongly
model dependent ~\cite{mv87},
{}~\cite{dr94}. Moreover it was shown recently that the cosmological
bounds can be evaded in perfectly reasonable scenarios of matter
genesis \cite{dr94}.

The effect of "soft" supersymmetry breaking can be parametrized
at the Fermi scale as a part of the scalar potential:

\ba{V_soft}
V_{soft}= \sum_{i=scalars}^{}  m_{i}^{2} |\phi_i|^2 +
h_L A_L H_1 \tilde L \tilde{\bar E} + h_D A_D H_1 \tilde Q \tilde{\bar D}
- h_U A_U H_2 \tilde Q \tilde{\bar U}  -&& \\
\nn
- \mu B H_1 H _2 + \mbox{ h.c.}&&
\ea

and a "soft"  gaugino mass term
\be{M_soft}
{\cal L}_{GM}\  = \ - \frac{1}{2}\left[M_{1}^{} \tilde B \tilde B +
 M_{2}^{} \tilde W^k \tilde W^k  + M_{3}^{} \tilde g^a \tilde g^a\right]
 -   \mbox{ h.c.}
\ee
As usual, $M_{3,2,1}$ are the masses of the $SU(3)\times
SU(2)\times U(1)$ gauginos $\tilde g, \tilde W, \tilde B$ and $m_i$
are the masses of scalar fields. $A_L,\ A_D, \ A_U$ and $B$ are
trilinear and bilinear "soft" supersymmetry breaking parameters.
Here, $m_{\tilde g} = M_3$ stands for the gluino $\tilde g$ mass.

Now the model is completely specified and we can deduce
the interaction terms of the  \rp  MSSM - Lagrangian  relevant for
neutrinoless double beta decay.

Let us write down these  interaction terms explicitly. In the
following  we use the 4-component Dirac bispinor notation
for fermion fields.

The lepton number violating part of the Lagrangian can be obtained directly
from the $W_{\rpm}$ superpotential part \rf{R-viol}. It  has the form
\ba{Lqqe}
        {\cal L}_{\rpm} = &-& \lambda'_{111}\left[
	(\bar{u}_L \ \bar{d}_R)\cdot
	\mbox{$ \left( \begin{array}{cc}
	e_{R}^{c}\\
	-\nu_{R}^{c}
 	\end{array} \right) $}\ \tilde{d}_R
	+
	(\bar{e}_L\ \bar{\nu}_L)\ d_R\cdot
	\mbox{$ \left( \begin{array}{cc}
	\tilde{u}_{L}^{\ast}\\
	-\tilde{d}_{L}^{\ast}
        \end{array} \right) $} + \right. \\ \nn
	&+& \left. (\bar{u}_L\ \bar{d}_L)\ d_R \cdot
	\mbox{$ \left( \begin{array}{cc}
	\tilde{e}_{L}^{\ast}\\
	-\tilde{\nu}_{L}^{\ast}
 	\end{array} \right) $}
	 + h.c. \right]
	\ea
To construct diagrams contributing to $\znbb$ decay we will also need
$R_p$ conserving gaugino-fermion-sfermion vertices.

The  Lagrangian terms corresponding to  gluino ${\cal L}_{\tilde{g}}$
and neutralinos ${\cal L}_{\chi}$ interactions with fermions
$\psi = \{u, d, e\}$, $q = \{u, d\}$ and their superpartners
$\tilde \psi = \{\tilde u, \tilde d, \tilde e\}$,
$\tilde q =\{\tilde u, \tilde d\}$
are \cite{Haber}
 \ba{gluino}
                {\cal L}_{\tilde g} = - \sqrt{2} g_3
        \frac{{\bf \lambda}^{(a)}_{\alpha \beta}}{2}
        \left( \bar q_L^{\alpha} \tilde g \tilde q_L^{\beta}
                          - \bar q_R^{\alpha} \tilde g \tilde q_R^{\beta}
                        \right)
			+ h.c.,
\ea
\ba{neutralino}
 	{\cal L}_{\chi} = \sqrt{2} g_2 \sum_{i=1}^4 \left(
                        \epsilon_{L i}(\psi) \bar \psi_L
				\chi_i \tilde \psi_L
        +  \epsilon_{R i}(\psi) \bar \psi_R \chi_i \tilde \psi_R\right)
			+ h.c.
\ea
Here
${\bf \lambda}^{(a)}$ are $3\times 3$ Gell-Mann matrices ($a = 1,..., 8$).
Neutralino coupling constants are defined as \cite{Haber}
\ba{eps}
        \epsilon_{L i}(\psi) &=& - T_3(\psi) {\cal N}_{i2} +
                                \tan \theta_W \left(T_3(\psi)
                        -  Q(\psi)\right) {\cal N}_{i1},\\ \nn
        \epsilon_{R i}(\psi) &=& Q(\psi) \tan \theta_W {\cal N}_{i1}.
\ea

Here $Q(\psi) $ and $ \ T_3(\psi)$ are the electric charge and weak
isospin of the field $\psi$.

The coefficients $N_{ij}$ are elements of the orthogonal
mixing matrix which diagonalizes the $4\times 4$ neutralino mass
matrix ~\cite{Haber}.
The four neutralino mass eigenstates $\chi_i$ with masses $m_{\chi_i}$
\footnote{The neutralino masses $m_{\chi_i}$ can be either
positive or negative if the orthogonal matrix $N$ is
used in the diagonalization
procedure ~\cite{Gunion}. The sign of the mass coincides with the CP-parity
of the corresponding neutralinos $\chi_i$.}
have the field content
\be{admix}
\chi_i = {\cal N}_{i1} \tilde{B} +  {\cal N}_{i2}  \tilde{W}^{3} +
{\cal N}_{i3} \tilde{H}_{1}^{0} + {\cal N}_{i4} \tilde{H}_{2}^{0}.
\ee
Recall again that we use notations
$\tilde{W}^{3}$, $\tilde{B}$ for neutral $SU(2)_L \times U(1)$
gauginos and  $\tilde{H}_{2}^{0}$, $\tilde{H}_{1}^{0}$
for higgsinos which are the superpartners of the two neutral Higgs boson
fields $H_1^0$ and $H_2^0$.

Having specified the Lagrangian interaction terms \rf{Lqqe} - \rf{neutralino}
one can construct diagrams describing the \rp MSSM contribution
to the $\znbb$ decay.
The complete set of these diagrams is presented in fig. 2.

The supersymmetric mechanism of $\znbb$ decay was first proposed
by Mohapatra \cite{Mohapatra}
and later studied in more details by Vergados \cite{Vergados1}.
In these papers \cite{Mohapatra}, \cite{Vergados1} only three diagrams
similar to those in Fig.2(a) were considered.
Instead of neutralinos $\chi_i$, which are actual mass eigenstates
in the MSSM, the consideration of refs. \cite{Mohapatra},~\cite{Vergados1}
used Z-ino ($\tilde Z$) and photino ($\tilde \gamma$) fields
in intermediate states. $\tilde Z$ and $\tilde \gamma$
can be mass eigenstates only at special values of parameters of
the neutralino mass matrix.
In general, these fields are not mass eigenstates. Using such fields
in the intermediate states leads to neglecting diagrams with mixed
intermediate states when, for instance,
$\tilde Z$ turns to $\tilde \gamma$ due to the mixing
proportional to the relevant entry of the neutralino mass matrix.
The effect of mixing is taken into account completely in the diagrams
displayed in fig.2(a,b) with all neutralino mass eigenstates $\chi_i$ involved.
Note that diagrams in fig. 2(b) were previously not considered
in the literature.

In the case of $0\nu\beta\beta$ decay when momenta of external particles
are much  smaller then intermediate particle masses one can treat the
interactions in Figs 1,2 as point-like.  A suitable formalism in
this case is the effective Lagrangian approach.

It is now straightforward to find the operators in the effective Lagrangian
which correspond to the diagrams in Fig. 2(a,b). The result is
\ba{Leff}
&&{\cal L}^{\Delta L_e =2}_{eff}(x)\ =\
8 \pi \alpha_2 \lambda^{'2}_{111} \sum_{i=1}^{4} \frac{1}{m_{\chi_i}}\ \left[
\frac{\epsilon_{L i}^2(e)}{m_{\tilde e_L}^4}
(\bar u_L^{\alpha} d_{R \alpha})(\bar u_L^{\beta} d_{R \beta})
(\bar e_L e^{\bf c}_{R}) + \right. \\ \nn
&&+ \left.\frac{\epsilon_{L i}^2(u)}{m_{\tilde u_L}^4}
(\bar u_L^{\alpha} u^{\bf c}_{R \beta})(\bar e_L d_{R \alpha})
(\bar e_L d_{R}^{\beta}) +
\frac{\epsilon_{R i}^2(d)}{m_{\tilde d_R}^4}
(\bar u_L^{\alpha} e^{\bf c}_R)(\bar u_L^{\beta} e^{\bf c}_R)
(\overline{ d^{\bf c}}_{L \alpha} d_{R \beta}) + \right.\\ \nn
&+& \left. \left(
\frac{\epsilon_{L i}(u)\epsilon_{R i}(d)}{m_{\tilde u_L}^2 m_{\tilde d_R}^2}
+ \frac{\epsilon_{L i}(u)\epsilon_{L i}(e)}{m_{\tilde u_L}^2 m_{\tilde e_L}^2}
+\frac{\epsilon_{L i}(e)\epsilon_{R i}(d)}{m_{\tilde e_L}^2 m_{\tilde d_R}^2}
\right) (\bar u_L^{\alpha} d_R^{\beta})(\bar u_{L \beta} e^{\bf c}_R)
(\overline{e}_L d_{R \alpha})
\right]\ + \\ \nn
&&+\  \lambda^{'2}_{111} \frac{8 \pi \alpha_s}{{m_{\tilde g}}}
\frac{{\bf \lambda}^{(a)}_{\alpha \beta}}{2}
\frac{{\bf \lambda}^{(a)}_{\gamma \delta}}{2}
 \left[
\frac{1}{m_{\tilde u_L}^4}
(\bar u_L^{\alpha} u^{{\bf c} \gamma} _{R})(\bar e_L d_R^{\beta})
(\bar e_L d_R^{\delta}) + \right.\\ \nn
&+& \left.\frac{1}{m_{\tilde d_R}^4}
(\bar u_L^{\alpha} e^{\bf c}_R)(\bar u_L^{\gamma} e^{\bf c}_R)
(\overline{ d^{\bf c}}_L^{\beta} d^{\bf c\ \delta}_{R})
- \frac{1}{m_{\tilde d_R}^2 m_{\tilde u_L}^2}
(\bar u_L^{\alpha} d_R^{\delta})(\bar u_L^{\gamma} e^{\bf c}_R)
(\overline{e}_L d_{R}^{\beta})
\right]\  \
\ea
Here $\alpha_2 = g_{2}^{2}/(4\pi)$ and $\alpha_s = g_{3}^{2}/(4\pi)$.

The Lagrangian \rf{Leff} has terms in a form which do not
allow a direct application of the non-relativistic impulse approximation
which is necessary for the standard calculation of the $\znbb$ reaction
matrix element ~\cite{hax84},~\cite{doi85}.
One should rearrange the right hand side of eq. \rf{Leff}  in
the form of a product
of two colour-singlet quark currents and the leptonic current.
It can be accomplished by a Fierz rearrangement procedure and subsequent
extraction of color-singlets from the product of two colour-triplet
and colour-antitriplet  quark fields.
The final result is
\ba{Leta}
{\cal L}^{\Delta L_e =2}_{eff}(x)\ &=&\
\frac{G_F^2}{2}\cdot m_P^{-1}\left[(\eta_{\tilde g} + \eta_{\chi})
(J_{PS}J_{PS} - \frac{1}{4} J_T^{\mu\nu} J_{T \mu\nu}) \ + \right. \\ \nn
&+& \left. (\eta_{\chi \tilde e} + \eta'_{\tilde g} - \eta_{\chi \tilde f})
J_{PS} J_{PS} + \eta_N J_{VA}^{\mu}J_{VA \mu}\right]
(\bar e (1 + \gamma_5) e^{\bf c}).
\ea
The last term, corresponding to the heavy Majorana neutrino exchange diagram
in fig.1, is included for completeness.
The lepton number violating parameters are defined as follows
\ba{eta}
\eta_{\tilde g} &=& \frac{2 \pi \alpha_s}{9}
\frac{\lambda^{'2}_{111}}{G_F^2 m_{\tilde d_R}^4} \frac{m_P}{m_{\tilde g}}
\left[
1 + \left(\frac{m_{\tilde d_R}}{m_{\tilde u_L}}\right)^4\right]\\
\eta_{\chi} &=& \frac{ \pi \alpha_2}{6}
\frac{\lambda^{'2}_{111}}{G_F^2 m_{\tilde d_R}^4} \sum_{i=1}^{4}
\frac{m_P}{m_{\chi_i}}
\left[
\epsilon_{R i}^2(d) + \epsilon_{L i}^2(u)
\left(\frac{m_{\tilde d_R}}{m_{\tilde u_L}}\right)^4\right]\\
\eta_{\chi \tilde e} &=& 2 \pi \alpha_2
\frac{\lambda^{'2}_{111}}{G_F^2 m_{\tilde d_R}^4}
\left(\frac{m_{\tilde d_R}}{m_{\tilde e_L}}\right)^4
\sum_{i=1}^{4}\epsilon_{L i}^2(e)\frac{m_P}{m_{\chi_i}},\\
\eta'_{\tilde g} &=& \frac{4 \pi \alpha_s}{9}
\frac{\lambda^{'2}_{111}}{G_F^2 m_{\tilde d_R}^4}
\frac{m_P}{m_{\tilde g}} \left(\frac{m_{\tilde d_R}}
{m_{\tilde u_L}}\right)^2,\\
\label{(eta_end)}
\eta_{\chi \tilde f} &=& \frac{\pi \alpha_2 }{3}
\frac{\lambda^{'2}_{111}}{G_F^2 m_{\tilde d_R}^4}
\left(\frac{m_{\tilde d_R}}{m_{\tilde e_L}}\right)^2
\sum_{i=1}^{4}\frac{m_P}{m_{\chi_i}}
\left[\epsilon_{R i}(d) \epsilon_{L i}(e)  + \right.\\ \nn
&+& \left.\epsilon_{L i}(u) \epsilon_{R i}(d)
\left(\frac{m_{\tilde e_L}}{m_{\tilde u_L}}\right)^2
+ \epsilon_{L i}(u) \epsilon_{L i}(e)
\left(\frac{m_{\tilde d_R}}{m_{\tilde u_L}}\right)^2
\right],\\
\label{(m_N)}
\eta_N &=& \frac{m_P}{<m_N>},
\ea

where $<m_N>$ is the {\em effective} heavy Majorana neutrino mass
(for definition see ~\cite{doi85}).

Colour-singlet hadronic currents have the form
\ba{Hcurr}
J_{PS} &=& \bar u^{\alpha} (1 + \gamma_5) d_{\alpha}, \
J_T^{\mu \nu} = \bar u^{\alpha} \sigma^{\mu \nu}(1 + \gamma_5) d_{\alpha},\\
J_{AV}^{\mu} &=& \bar u^{\alpha} \gamma^{\mu}(1 - \gamma_5) d_{\alpha},
\label{(JT)}
\ea
where $\sigma^{\mu \nu} = \frac{i}{2} [\gamma^{\mu}, \gamma^{\nu}]$.

Let us write down the $\znbb$ decay matrix element ${\cal R}_{\znbb}$
corresponding to  the effective Lagrangian eq. \rf{Leta}
\ba{R_0nu}
{\cal R}_{\znbb} &=& \frac{G_{F}^{2}}{\sqrt{2}} m_{P}^{-1} C_{0\nu}^{-1}
\left\{\bar{e} (1 + \gamma_5) e^{\bf c} \right\}\times \\ \nn
&&\left[
\left(\eta_{\tilde g} + \eta_{\chi} \right)\Big<F| \Omega_{\tilde q}
|I\Big> +
\left(\eta_{\chi \tilde e} + \eta'_{\tilde g} - \eta_{\chi \tilde f}
\right)\Big<F| \Omega_{\tilde f} |I \Big>
+ \eta_N \Big<F| \Omega_N |I \Big> \right].
\ea
The heavy Majorana neutrino exchange contribution has been included
in the last term of this equation. It corresponds to the last term
of eq. \rf{Leta}. The normalization factor $C_{0\nu}$ is defined below.

We have introduced  transition operators  $\Omega_i$
(for definitions see ~\cite{doi85}) which are useful for separating
the particle physics part of the calculation from
the nuclear physics one. The transition operators contain information
about the underlying interactions at the quark level \rf{Leta}
and quark states
inside the nucleon. They  are independent of the initial $|I>$ and
the final $<F|$ nuclear states. To calculate the nuclear matrix elements
in eq. \rf{R_0nu} one may use any nuclear model wave functions.
In our case $\Omega_{\tilde q}, \Omega_{\tilde f}$ and
$\Omega_N$ describe
transitions induced by quark colour singlet currents \rf{Hcurr}-\rf{JT}
in the first, second and third terms of eq. \rf{Leta}. Diagrams in fig.1,2
with the intermediate states $\{W - N - W\}$,
$\{\tilde u(\tilde d) - \chi,\tilde g - \tilde u(\tilde d)\}$ and
$\{\tilde u(\tilde d) - \chi,\tilde g - \tilde d(\tilde u);
\tilde e - \chi - \tilde e,\tilde q\}$
contribute to operators $\Omega_N$, $\Omega_{\tilde q}$ and
$\Omega_{\tilde f}$ respectively.
Calculation of the transition operators in the non-relativistic
impulse approximation (NRIA) ~\cite{doi85} requires
the nucleon matrix elements of these currents.

We derive nucleon matrix elements of the colour singlet quark  currents
\rf{Hcurr}-\rf{JT} using results of ref. \cite{Adler}.
The necessary matrix elements are,
\ba{PS}
<P(p)| \bar u d|N(p')> &=& F_S^{(3)}(q^2)\cdot \bar N(p) \tau_{+} N(p'),\\
<P(p)| \bar u \gamma_5 d|N(p')> &=& F_P^{(3)}(q^2)\cdot \bar N(p)
				    \gamma_5\tau_{+} N(p'),\\
<P(p)| \bar u \sigma^{\mu \nu}(1 + \gamma_5) d|N(p')>
	&=& \bar N(p) \left(J^{\mu\nu} +
\frac{i}{2} \epsilon^{\mu\nu\rho\sigma} J_{\rho\sigma}\right)
\tau_{+} N(p'),\\
<P(p)|\bar u \gamma^{\mu}(1 - \gamma_5) d |N(p')>
&=& \bar N(p)\gamma^{\mu}\left(F_V(q^2) - F_A(q^2)\gamma_5\right)\tau_{+}
N(p'),
\ea
where $q = p - p'$ and the tensor structure is defined as
\ba{JJJ}
J^{\mu\nu} = T_1^{(3)}(q^2) \sigma^{\mu\nu} + \frac{i T_2^{(3)}}{m_P}
\left(\gamma^{\mu} q^{\nu} -  \gamma^{\nu} q^{\mu}\right) +
\frac{T_3^{(3)}}{m_P^2}
\left(\sigma^{\mu\rho} q_{\rho} q^{\nu} -  \sigma^{\nu\rho} q_{\rho}
q^{\mu}\right).
\ea
For all form factors $F_{V,A}(q^2), F_{S,P}^{(3)}(q^2),
T_i^{(3)}(q^2)$ we take,
following ref. \cite{Vergados2}, a dipole form
\ba{dip}
\frac{F_{V,A}(q^2)}{f_{V,A}} =
\frac{F_{S,P}^{(3)}(q^2)}{F_{S,P}^{(3)}(0)} =
\frac{T_i^{(3)}(q^2)}{T_i^{(3)}(0)} =
\left(1 - \frac{q^2}{m_A^2} \right)^{-2}
\ea
with $m_A = 0.85$GeV and $f_V\approx 1,\ f_A\approx 1.261$.
Form factor normalization values $F_{S,P}^{(3)}(0)$,
$T_i^{(3)}(0)$
were calculated in ref. \cite{Adler} and are given in Table 1.

\begin{table}[t]
{Table 1: Nucleon form factor normalizations at $q^2 =0$ as calculated in
\cite{Adler}.}\\[5mm]
\begin{tabular}{|c|c|c|c|c|c|}\hline
Set & $F_S^{(3)}$ & $F_P^{(3)}$ & $T_1^{(3)}$ & $T_2^{(3)}$ & $T_3^{(3)}$\\
\hline
A) Bag Model  & 0.48 & 4.41 & 1.38 & -3.30 & -0.62\\
\hline
B) Non-relativistic  & 0.62 & 4.65 & 1.45 & -1.48 & -0.66\\
quark model&&&&&\\
\hline
\end{tabular}
\end{table}

Using formulas \rf{PS}-\rf{JJJ} we may derive the non-relativistic limit
$m_P \gg |\vec p|$ for
nucleon matrix elements of the three combinations of the quark currents
in \rf{Leta}. Keeping all terms up to order $q^2$ in
the non-relativistic expansion we find
the relevant transition operators
$\Omega_{\tilde q}, \Omega_{\tilde f}, \Omega_N$ for the case of
two outgoing electrons in S-wave states,
\ba{Omega}
\Omega_{\tilde q} &=&
{\frac{m_P}{ m_e}}\Big\{ \alpha_V^{(0)} \Omega_{F,N} +
\alpha_A^{(0)}\Omega_{GT,N} + \alpha_V^{(1)} \Omega_{F'}  +
\alpha_A^{(1)} \Omega_{GT'} + \alpha_{T} \Omega_{T'} \Big\},\\
\label{(Omega_e)}
\Omega_{\tilde f} &=& \Omega_{\tilde q}(T_i = 0),\\
\label{(Omega_N)}
\Omega_N &=&
{\frac{m_P}{ m_e}}\Big\{ \left(\frac{f_V}{f_A}\right)^2 \Omega_{F,N} -
                        \Omega_{GT,N} \Big\},
\ea
where partial transition operators are
\ba{O_i}
\Omega_{GT,N} &=&  \sum_{i \ne j} \tau_{+}^{(i)} \tau_{+}^{(j)}
                     \si \cdot \sj
                     \left(\frac{R_0}{r_{ij}}\right)
                      F_{N}(x_{A}),\\
\Omega_{F,N} &=& \sum_{i \ne j} \tau_{+}^{(i)} \tau_{+}^{(j)}
                     \left(\frac{R_0}{r_{ij}}\right) F_{N}(x_{A}),\\
\label{(GTT)}
\Omega_{GT'} &=&   \sum_{i \ne j} \tau_{+}^{(i)} \tau_{+}^{(j)}
                     \si \cdot \sj
                     \left(\frac{R_0}{r_{ij}}\right) F_{4}(x_{A}),\\
\Omega_{F'}  &=& \sum_{i \ne j} \tau_{+}^{(i)} \tau_{+}^{(j)}
                 \left(\frac{R_0}{r_{ij}}\right) F_{4}(x_{A}),\\
\label{(O_T)}
\Omega_{T'} &=& \sum_{i \ne j} \tau_{+}^{(i)} \tau_{+}^{(j)}
                     \big\{ 3 \sir \sjr -
                     \si \cdot \sj \big\}
                     \left(\frac{R_0}{r_{ij}}\right) F_{5}(x_{A}).
\ea
Here, $R_0$ is the nuclear radius, introduced to make the matrix
elements dimensionless (compensating factors have been absorbed into
the phase space integrals \cite{doi85}).
The following notations are used
${\bfr}_{ij} = ({\overrightarrow r}_i -
                         {\overrightarrow r}_j ),\
r_{ij} = |{\bfr}_{ij}|, \  {\hat{\bfr}_{ij}^{~}}= {\bfr}_{ij}/r_{ij}, \
x_A = m_A r_{ij}$. We also define the normalization constant in eq. \rf{R_0nu}
$ C_{0\nu} = 4 \pi \frac{m_P}{m_e} \frac{R_0}{f_{A}^{2}} m_A^{-2}$.
The above formulae have been written in the closure approximation, which
is well satisfied for large masses of intermediate
particles ~\cite{doi85}.

The nucleon structure coefficients in \rf{Omega} are given by
\ba{alph}
\alpha^{(0)}_V &=& \left(\frac{ F_S^{(3)}}{f_A}\right)^2,
\alpha^{(0)}_A = -\left(\frac{ T_1^{(3)}}{f_A}\right)^2,\\
\alpha^{(1)}_V &=& -\left(\frac{m_A}{m_P}\right)^2 \alpha^{(0)}_A
\left[\frac{1}{4} + \left(\frac{T_2^{(3)}}{T_1^{(3)}}\right)^2 -
\frac{T_2^{(3)}}{T_1^{(3)}}\right],\\
\alpha^{(1)}_A &=& -\left(\frac{m_A}{m_P}\right)^2 \left[\alpha^{(0)}_A
\left(\frac{1}{6} - \frac{2}{3} \frac{T_2^{(3)}}{T_1^{(3)}} +
\frac{4}{3} \frac{T_3^{(3)}}{T_1^{(3)}}\right) +
\frac{1}{12} \left(\frac{F_P^{(3)}}{f_A}\right)^2\right],\\
\alpha_T &=& -\left(\frac{m_A}{m_P}\right)^2 \left[\alpha^{(0)}_A
\left(\frac{1}{12} - \frac{1}{3} \frac{T_2^{(3)}}{T_1^{(3)}} + \frac{2}{3}
\frac{T_3^{(3)}}{T_1^{(3)}}\right) - \frac{1}{12}
\left(\frac{F_P^{(3)}}{f_A}\right)^2\right].
\label{(alpha_end)}
\ea
Here
$F_{S,P}^{(3)} \equiv F_{S,P}^{(3)}(0),\ T_{i}^{(3)} \equiv T_{i}^{(3)}(0)$.
Three different structure functions $F_i$ appear in eq. \rf{Omega}
(we use notations of ref. ~\cite{Vergados1}).
They are given by
\ba{Fnn}
&F_N(x)  = \frac{x}{48} (3 + 3 x + x^2) e^{-x}, \ \
F_4(x)  = \frac{x}{48} (3 + 3 x - x^2) e^{-x},&\\ \nn
&F_5(x)  = \frac{x^3}{48} e^{-x}.&
\ea
These functions are the analytic solutions of
the relevant integrals over the intermediate particle momentum.
They are analogons to the "neutrino potentials" for the
case of light Majorana neutrino exchange ~\cite{doi85}.

At this stage we  point out that our formulas for
the coefficients \rf{alph}-\rf{alpha_end} of the  transition
operators \rf{Omega}-\rf{Omega_e} disagree with the corresponding
formulas derived in ref. \cite{Vergados1}. Particularly, in our case
$\alpha_A^{(0)} \leq 0$ while in ref. \cite{Vergados1}
this coefficient is positive. This sign difference has
an important consequence if one considers simultaneously
the supersymmetric and the heavy Majorana neutrino contributions.
Our formulas correspond to a constructive interference between
these two mechanisms while formulas from ref. \cite{Vergados1}
correspond to a destructive one.
In the latter case both contributions can
cancel each other and by a proper choice of $\lambda'_{111}$
in eqs. \rf{eta}-\rf{eta_end}
the  matrix element of $\znbb$-decay  can be set to zero at any values
of particle masses involved in the formulas. As a result the $\znbb$-decay
half-life limit would neither constrain supersymmetric
particle masses nor that for Majorana neutrinos.
Our formulas \rf{Omega}-\rf{alpha_end} always lead to certain
constraints on these masses. Detailed discussion of this point
will be given elsewhere.

 In conclusion we note that the results obtained in the present paper
can be directly used for the extraction of constraints on the supersymmetric
parameter space from the non-observation of $\znbb$ decay. To do this
one should employ nuclear wave functions calculated in a specific
nuclear structure model. A subsequent paper will be devoted
to this subject.

\bigskip
\centerline{\bf ACKNOWLEDGMENTS}

We thank V.A.~Bednyakov, V.B.~Brudanin,
M.~Lindner and
J.W.F.~Valle  for helpful discussions.
The research described in this publication was made possible in part by
Grant No.RFM000 from the international Science Foundation.

{\large\bf Figure Captions}\\

\begin{itemize}

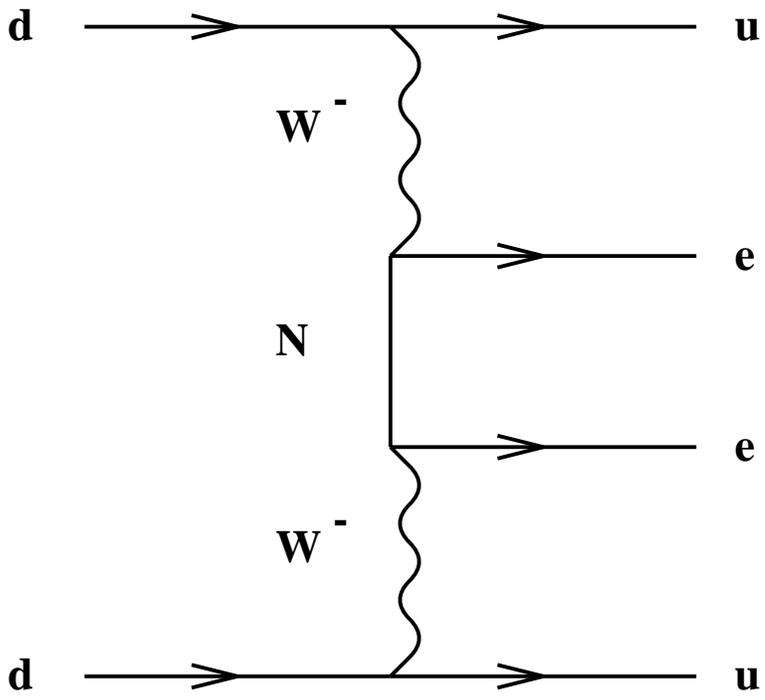
\item[Fig.1] Feynman graphs for the conventional mechanism
of $\znbb$ decay by  massive Majorana neutrino exchange.

\item[Fig.2a] Feynman graphs for  supersymmetric contributions
to $\znbb$ decay.

\item[Fig.2b] New "non-diagonal" Feynman graphs for supersymmetric
contributions to $\znbb$ decay.
\end{itemize}
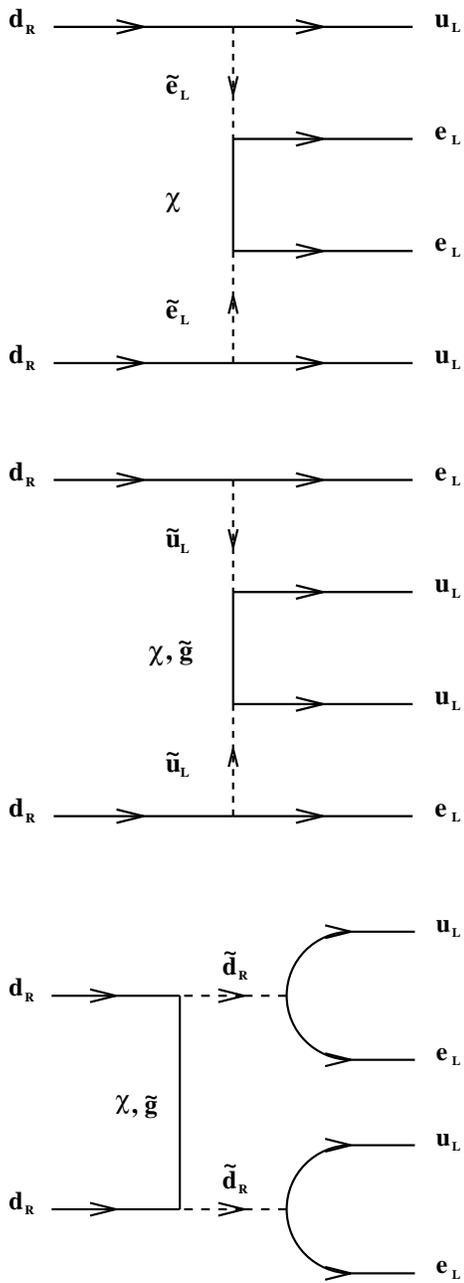
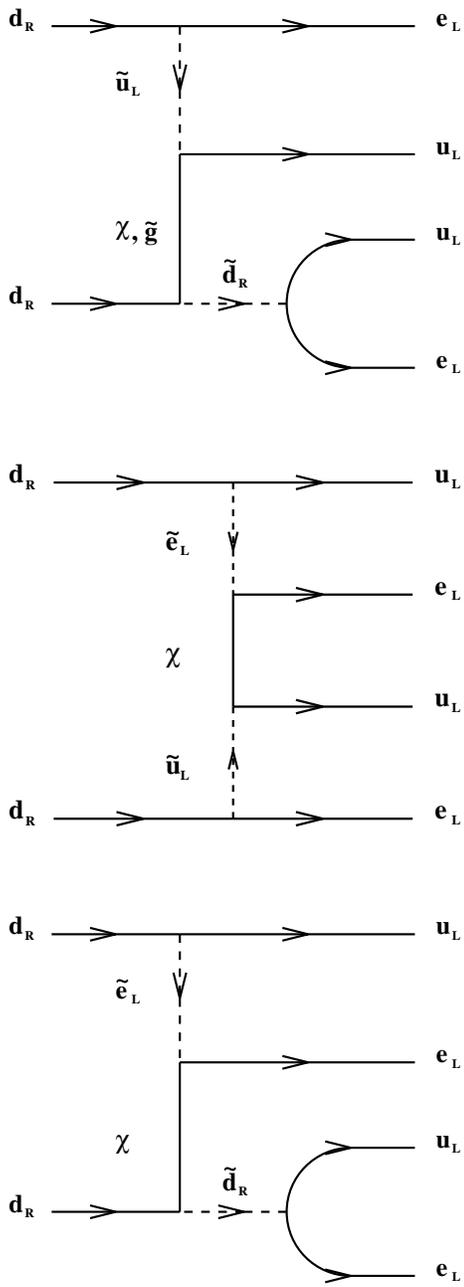
\end{document}